\documentclass[onecolumn,showpacs,aps,pre,amsmath,amssymb,tightenlines,11pt]{revtex4}
\usepackage{graphicx}% Include figure files
\usepackage{dcolumn}% Align table columns on decimal point
\usepackage{bm}% bold math

\usepackage{amsmath}

\def\be{\begin{equation}}
\def\ee{\end{equation}}
\def\bea{\begin{eqnarray}}
\def\eea{\end{eqnarray}}
\def\ba{\begin{array}}
\def\ea{\end{array}}

\def\0{$\Gamma_0$}

\begin{document}

\title{Correlations among the Riemann zeros: Invariance, resurgence, prophecy and
self-duality}
\author{W. T. Lu$^{\dagger }$ and S. Sridhar*}
\affiliation{Department of Physics and Electronic Materials Research
Institute, Northeastern University, Boston, Massachusetts 02115}
\date{\today }

\begin{abstract}
We present a conjecture describing new long range correlations among the
Riemann zeros leading to 3 principal features:\ (i) The spectral
auto-correlation is \emph{invariant} w.r.t. the averaging window. (ii) \emph{%
Resurgence} occurs wherein the lowest zeros appear in all auto-correlations.
(iii) Suitably defined correlations lead to predictions (\emph{prophecy}) of
new zeros. This conjecture is supported by analytical arguments and
confirmed by numerical calculations using $10^{22}$ zeros computed by
Odlyzko. The results lead to a self-duality of the Riemann spectrum similar
to the quantum-classical duality observed in billiards.
\end{abstract}


\maketitle

\section{Introduction}

The Riemann zeta function is 
\[
\zeta (s)=\sum\limits_{n=1}^{\infty }\frac{1}{n^{s}}=\prod%
\limits_{p}(1-p^{-s})^{-1}
\]
with product over all prime numbers $p$. The Riemann zeta function has a
simple pole at $s=1$. Besides the trivial zeros at $s=-2,-4,-6,\cdots $, the
Riemann hypothesis (RH) states that all the nontrivial zeros are distributed
on line as $\frac{1}{2}\pm it_{n}$ with $t_{n}>0$ \cite{Conrey}. The Riemann
zeta function and its zeros have been studied for over a century \cite
{Edwards,Titch,Berry140,Berry175,Keating96,Berry306,Berry307,Keating,Connors,Odly,Bohigas01,Leboeuf01,Zyl,Sakhr03,Bogomolny03,Leboeuf04}. 
The RH still remains unproven and presents a major mathematical challenge 
\cite{Sabbagh}.

Motivated by the Hilbert-Polya conjecture of the existence of a Hamiltonian
whose quantum spectrum is comprised of the set $\{t_{n}\}$ as the spectra,
there has been a continuing quest to find such a Hamiltonian since it would
prove the RH. Attempts have been made to construct certain Hamiltonian such
as that with a fractal potential \cite{Wu93} for the Riemann zeros (RZ). For
a brief review about Hamiltonians and prime numbers, see \cite{Rosu}.

The RZ $\{\frac{1}{2}\pm it_{n}\}$ can also be considered as a model
spectrum for arithmetic or quantum chaos \cite{Sarnak}. It is widely
believed that the ordinates $\{t_{n}\}$ of the RZ can be interpreted as the
quantum spectrum of a chaotic system which does not have time reversal
symmetry \cite{Berry154,Berry307} though other interpretations are also
possible. The most notable result in support of this notion is that of
Montgomery \cite{Montgomery} who showed that the spectral correlation
function $R_{2}(s)$ of the RZ is identical with that of a Gaussian Unitary
Ensemble (GUE) which describes a chaotic system with no time-reversal
symmetry in random matrix theory. This result has led to a line of inquiry
which studies the statistics and correlations among the RZ. The merit of
this approach is that the properties of the RZ will lead to insights into
other systems where the classical dynamics is well established. These are
motivated by the analog of the Riemann zeta function with Gutzwiller's trace
formula and the dynamic zeta function of chaotic systems.

In this paper, we present a conjecture which reveals new long range
correlations among the RZ, obtained from a study of the auto-correlation of
a smoothed Riemann spectrum. Several new properties emerge:\ (i) The
spectral auto-correlation is \emph{invariant} w.r.t. the averaging window.
We demonstrate numerically that this remarkable invariance applies over any
part of the spectrum of $10^{22}$ zeros, and clearly suggests that the same
structure is encoded in all parts of the spectrum. (ii) \emph{Resurgence}
occurs wherein the lowest zeros appear in all auto-correlations. This
property of resurgence was first pointed out by Berry and Keating \cite
{Berry307}. (iii) Suitably defined correlations lead to predictions (\emph{%
prophecy}) of new zeros.

The property of resurgence and prophecy suggests a \emph{self-duality }for
the RZ which we represent as $C\{\frac{1}{2}\pm it_{n}\}\Rightarrow $ $\{%
\frac{1}{2}\pm it_{n}\}$. This is very similar to a duality $%
C\{k_{n}+ik_{n}^{\prime }\}\Rightarrow $ $\{\gamma _{i}^{^{\prime }}+i\gamma
_{i}^{\prime \prime }\}$ which is observed from the auto-correlation of the
quantum spectrum of open chaotic billiards, and which relates the quantum
resonance spectrum $\{k_{n}+ik_{n}^{\prime }\}$ and classical resonance
spectrum $\{\gamma _{i}^{^{\prime }}+i\gamma _{i}^{\prime \prime }\}$ of the
hyperbolic $n$-disk repeller \cite{Pance}.

The paper is organized in the following. In Sec. II, we introduce a smoothed
spectral function and the spectral correlations and describe a conjecture
based upon the spectral correlations. A ``semiclassical''\ derivation of the
conjecture is presented in Sec. III. Numerical calculations supporting the
properties of invariance, resurgence, prophecy and self-duality embodied in
the conjecture are presented in Sec. IV. In Sec. V, we discuss the
self-duality of the RZ with the quantum-classical duality observed in
several billiard systems, and discuss the implications for a dynamic
interpretation of the RZ.

\section{Correlations and a conjecture}

The spectral density is the sum of $\delta $-function $\rho
(k)=\sum_{n}\delta (k\pm t_{n})$. In order to study the correlation among
the RZ, we use a Lorentzian-smoothed spectral density

\[
\rho _{\epsilon }(k)=\frac{1}{\pi }\sum\limits_{n}\frac{\epsilon }{(k\pm
t_{n})^{2}+\epsilon ^{2}}.
\]
Here $t_{n}$ are the ordinates of the RZ and $\epsilon >0$ is a small width.
In the limit $\epsilon \rightarrow 0$, one gets the stick spectrum. It can
be written as a continuous part plus fluctuations 
$\rho _{\epsilon }(k)=\left\langle \rho _{\epsilon }(k)\right\rangle +\delta
\rho _{\epsilon }(k)$.
The continuous part of the spectral density for the nontrivial RZ \cite
{Berry140} is $\left\langle \rho (k)\right\rangle =(1/2\pi )\ln (k/2\pi )$
in the limit $\epsilon \rightarrow 0$. We define the following fluctuation
part of the spectral density \cite{Brack} 
\begin{eqnarray}
\delta \rho _{\epsilon }(k) &=&-\frac{1}{\pi }\frac{d}{dk}[\Im\ln \zeta (%
\frac{1}{2}+\epsilon +ik)]  \nonumber \\
&=&-\frac{1}{2\pi }\ln \frac{\sqrt{\epsilon ^{2}+k^{2}}}{2\pi }+\frac{1}{\pi 
}\sum\limits_{n=1}^{\infty }\left[ \frac{\epsilon }{(k+t_{n})^{2}+\epsilon
^{2}}+\frac{\epsilon }{(k-t_{n})^{2}+\epsilon ^{2}}\right] .  \label{dos-1}
\end{eqnarray}
The function $\delta \rho _{\epsilon }(k)$ defined above is an even
function. Note that the above quantity $\delta \rho _{\epsilon }(k)$ is
related to the fluctuation part of the quantum time delay $\tau _{\text{fl}%
}(w)$ considered in \cite{Wardlaw}, $\delta \rho _{1/2}(k)=\pi \tau _{\text{%
fl}}(k/2)/4$. In practice, $\delta \rho _{\epsilon }(k)$ is a finite sum.
For the spectral function $\delta \rho _{\epsilon }(k)$ in a window $k\in
[K_{0}-\Delta ,K_{0}+\Delta ]$ with $K_{0}\gg \Delta \gg 1$, one notices
that the contribution of RZ far outside $[K_{0}-\Delta ,K_{0}+\Delta ]$ is
negligible for small $\epsilon $. Thus one may use a finite sum 
\begin{equation}
\delta \rho _{\epsilon }(k)\simeq -\frac{1}{2\pi }\ln \frac{k}{2\pi }+\frac{1%
}{\pi }\sum\limits_{n=N_{0}}^{N}\frac{\epsilon }{(k-t_{n})^{2}+\epsilon ^{2}}%
.  \label{dos-2}
\end{equation}
The RZ used in the above expression should be slightly outside the window
such as $t_{N_{0}}\lesssim K_{0}-\Delta $ and $t_{N}\gtrsim K_{0}+\Delta $.

Define the following cross correlation of $\delta \rho _{\epsilon }(k)$
among two windows $[K_{1}-\Delta ,K_{1}+\Delta ]$ and $[K_{2}-\Delta
,K_{2}+\Delta ]$ with $K_{1}$, $K_{2}$ any real numbers and $K_{1}\leq K_{2}$
as 
\begin{eqnarray}
C_{\rho _{\epsilon }}^{c}(s;K_{1},K_{2},\Delta ) &\equiv &\{\left\langle
\delta \rho _{\epsilon }(K_{1}+k)\delta \rho _{\epsilon
}(K_{2}+k+s)\right\rangle _{k};\quad |s|\leq \Delta ,\text{\ }  \nonumber \\
&&\quad k\in \lbrack -\Delta -\min (0,s),\Delta -\max (0,s)]\}.
\label{cor-1}
\end{eqnarray}
The superscript \textquotedblleft $c$\textquotedblright\ denotes \textsl{%
cross}. Here the average over $k$ is defined as 
\[
\langle f_{1}(k)f_{2}(k)\rangle ={\frac{1}{b-a}}%
\int_{a}^{b}f_{1}(k)f_{2}(k)dk\,.
\]
There are two special cases 
\begin{eqnarray}
C_{\rho _{\epsilon }}^{r}(s;K_{0},\Delta ) &\equiv &C_{\rho _{\epsilon
}}^{c}(s;K_{0},K_{0},\Delta )\equiv \left\langle \delta \rho _{\epsilon
}(K_{0}+k)\delta \rho _{\epsilon }(K_{0}+k+s)\right\rangle _{k},
\label{cor-2} \\
C_{\rho _{\epsilon }}^{p}(s;K_{0},\Delta ) &\equiv &C_{\rho _{\epsilon
}}^{c}(s;K_{0},-K_{0},\Delta )\equiv \left\langle \delta \rho _{\epsilon
}(K_{0}-k)\delta \rho _{\epsilon }(K_{0}+k+s)\right\rangle _{k}.
\label{cor-3}
\end{eqnarray}
In Eq. (\ref{cor-3}), we used $\delta \rho _{\epsilon }(K_{0}-k)=\delta \rho
_{\epsilon }(-K_{0}+k)$. The superscript \textquotedblleft $r$%
\textquotedblright\ and \textquotedblleft $p$\textquotedblright\ denotes 
\emph{resurgence} and \emph{prophecy}, respectively. The meaning will be
apparent later.

We now summarize the principal results of this paper. We have the following

\emph{CONJECTURE}:

1. \emph{INVARIANCE} the spectral correlation is \emph{invariant and
independent of averaging window}. That is 
\begin{eqnarray}
C_{\rho _{\epsilon }}^{c}(s;K_{1},K_{2},\Delta ) &=&C^{\epsilon
}(K_{2}-K_{1}+s), \\
C_{\rho _{\epsilon }}^{r}(s;K_{0},\Delta ) &=&C^{\epsilon }(s), \\
C_{\rho _{\epsilon }}^{p}(s;K_{0},\Delta ) &=&C^{\epsilon }(2K_{0}+s)
\end{eqnarray}
Note that one has $-\Delta \leq s\leq \Delta $ with $\Delta \gg \,\langle
\delta =t_{n}-t_{n-1}\rangle $, the mean spacing between the RZ in a window.

2. The spectral correlation is a sum of functions of the pole and zeros of
the Riemann zeta function 
\begin{equation}
C^{\epsilon }(s)=\Re\left[ g(1+2\epsilon +is,1)-\sum\limits_{n=1}^{\infty
}g(1+2\epsilon +is,-2n)-\sum\limits_{n=1}^{\infty }g(1+2\epsilon +is,{\frac{1%
}{2}}\pm it_{n})\right] .  \label{conj-1}
\end{equation}
Here $g(z,\alpha )$ is a certain two-variable function. The second (third)
term is the summation over trivial (nontrivial) RZ.

3. The function $g(z,\alpha )$ can be broken into diagonal and off-diagonal
terms 
\begin{equation}
g(z,\alpha )=g_{\text{diag}}(z,\alpha )+g_{\text{off}}(z,\alpha ).
\end{equation}
Here $g_{\text{off}}\ll $ $g_{\text{diag}}$. Then we show that 
\begin{equation}
g_{\text{diag}}(z,\alpha )={\frac{1}{2\pi ^{2}}\frac{1}{(z-\alpha )^{2}}-%
\frac{1}{2\pi ^{2}}\sum\limits_{n=2}^{\infty }\frac{c_{n}}{(z-\alpha /n)^{2}}%
}  \label{g-diag}
\end{equation}
with $c_{n}$'s real coefficients, $|c_{n}|\leq 1/n$, and $c_{n}\rightarrow 0$
for $n\rightarrow \infty $. For the correlation function of interest $%
C_{\rho _{\epsilon }}(s)$ with $s\in [a,b]$, it can be replaced with a \emph{%
finite sum} over the pole and zeros of the Riemann\ zeta function 
\begin{eqnarray}
C^{\epsilon }(s)\simeq {\frac{1}{2\pi ^{2}}\Re}\left[ {\frac{1}{%
(is+2\epsilon )^{2}}-\sum\limits_{n=1}^{\infty }\frac{1}{(is+{1+}2\epsilon
+2n)^{2}}-\sum\limits_{n=N_{0}}^{N}\frac{1}{(is-it_{n}+{\frac{1}{2}+}%
2\epsilon )^{2}}}\right] , &&  \label{conjecture} \\
s\in [a,b],\quad t_{N_{0}}\lesssim a,\quad t_{N}\gtrsim b. &&  \nonumber
\end{eqnarray}
The contribution of RZ far outside $[a,b]$ is small.

4. \emph{RESURGENCE }

\begin{equation}
C_{\rho _{\epsilon }}^{r}(s;K_{0},\Delta )=C^{\epsilon }(s)\approx -{\frac{1%
}{2\pi ^{2}}}\Re{\sum\limits_{n=1}^{N}\frac{1}{(is\pm it_{n}+{\frac{1}{2}+}%
2\epsilon )^{2}}},\text{ }|s|<\Delta \text{, }t_{N}\gtrsim \Delta .
\label{conj-resurge}
\end{equation}

\begin{figure}[tbp]
\center{
\includegraphics [angle=0,width=9cm]{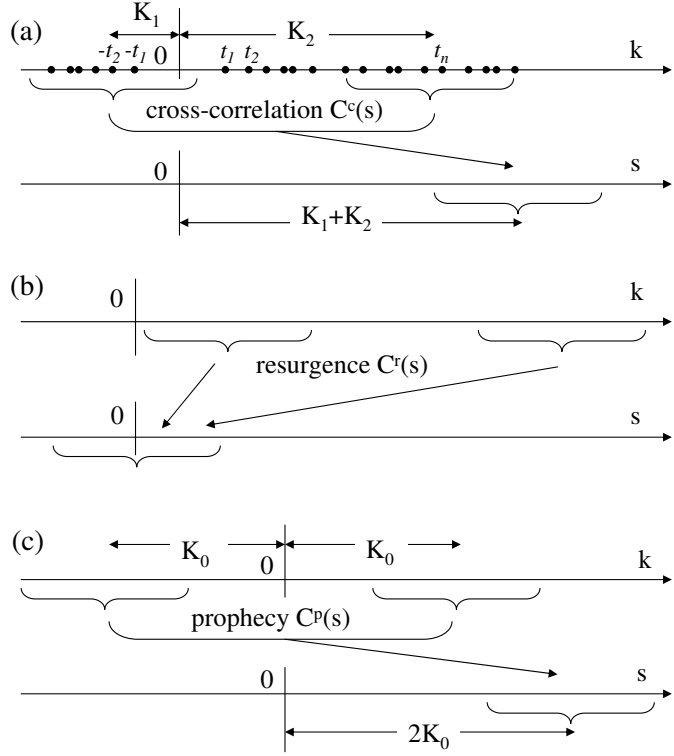}}
\caption{A sketch of the conjecture: (a) cross-correlation $C_{\rho
_{\epsilon }}^r(s)$, (b) invariance and resurgence in auto-correlation $%
C_{\rho _{\epsilon }}^r(s)$, (c) prophecy $C_{\rho _{\epsilon }}^p(s)$.}
\label{fig1}
\end{figure}

5. \emph{PROPHECY}

\begin{eqnarray}
C_{\rho _{\epsilon }}^{p}(s;K_{0},\Delta )=C^{\epsilon }(2K_{0}+s)\approx -{%
\frac{1}{2\pi ^{2}}}\Re{\sum\limits_{n=N_{0}}^{N}\frac{1}{(is+2iK_{0}-it_{n}+%
{\frac{1}{2}+}2\epsilon )^{2}}}, &&  \label{conj-prophecy} \\
|s|<\Delta \text{, }t_{N_{0}}\lesssim 2K_{0}-\Delta ,\quad t_{N}\gtrsim
2K_{0}+\Delta . &&  \nonumber
\end{eqnarray}
A graphic representation of the conjecture is shown in Fig.\ref{fig1}.

A few remarks are in order.

\emph{INVARIANCE }The\emph{\ }invariance of the correlations means that 
\emph{the same structure is encoded in the Riemann spectrum independent of
the center and width of the averaging window}. The cross correlation $%
C_{\rho _{\epsilon }}^{c}(s;K_{1},K_{2},\Delta )$ between the spectral
function $\delta \rho _{\epsilon }(k)$ in any two windows $[K_{1}-\Delta
,K_{1}+\Delta ]$ and $[K_{2}-\Delta ,K_{2}+\Delta ]$ is the same if the
center distance $K_{2}-K_{1}$ is the same.

\emph{RESURGENCE} We find that the peaks of $-C_{\rho _{\epsilon }}^{r}(s)$
are located at the positions $s=t_{n}$. The correlation between RZ in $%
[K_{0}-\Delta ,K_{0}+\Delta ]$ leads to RZ in $[-\Delta ,\Delta ]$. The
auto-correlation $C_{\rho _{\epsilon }}^{r}(s)$ in Eq. (\ref{cor-2}) is
essentially the two-level correlation $R_{2}(s)$ \cite{Keating96}.
Resurgence was first noted by Berry and Keating \cite{Berry307,Keating,Connors} 
and has been revisited recently by Leboeuf 
and collaborators \cite{Bohigas01,Leboeuf01,Leboeuf04}. While the
correlations studied in the present paper are related to the above work, the
difference is that here, there is no unfolding and no rescaling by the mean
level spacing, and reveals some unexpected new results among the RZ
summarized in the equations above.

\emph{PROPHECY} can lead to the prediction of new zeros in the interval $%
[2K_{0}-\Delta ,2K_{0}+\Delta ]$ from interval $[K_{0}-\Delta ,K_{0}+\Delta
] $.

Resurgence and prophecy can also be observed in the cross correlation $%
C_{\rho _{\epsilon }}^{c}(s;K_{1},K_{2},\Delta )$.

\section{Trace formula derivation}

The above properties are ``derived''\ using a semiclassical method \cite
{Agam95}. The fluctuation part of the spectral density of the RZ is \cite
{Berry140,Brack}

\begin{equation}
\delta \rho _{\epsilon }(k)=-{\frac{1}{\pi }}\sum_{p}L_{p}\sum_{r=1}^{\infty
}\Lambda _{p}^{-r({\frac{1}{2}}+\epsilon )}\cos (rkL_{p})
\end{equation}
Note that this is exactly of the same form as the Gutzwiller trace formula
for chaotic billiards in wave number $k$ \cite{Gutzwiller}, with length of
``periodic orbit''\ equal to the logarithm of prime numbers $L_{p}=\ln p$
and $\Lambda _{p}=e^{L_{p}}=p$. Like that of the chaotic systems such as the
chaotic $n$-disk systems \cite{Pance}, the number of ``periodic orbits''\
for the RZ grows asymptotically as $N(L_{p}<L)\sim e^{L}/L$. Although the
above trace formula converges only for $\epsilon >\frac{1}{2}$, we may still
use it formally for $\epsilon \rightarrow 0$. The correlation can be written
as the sum of diagonal and off-diagonal terms

\[
C_{\rho }^{\epsilon }(s)=\langle \delta \rho _{\epsilon }(k)\delta \rho
_{\epsilon }(k+s)\rangle =C_{\text{diag}}^{\epsilon }(s)+C_{\text{off}%
}^{\epsilon }(s).
\]
The diagonal part is 
\begin{equation}
C_{\text{diag}}^{\epsilon }(s)={\frac{1}{2\pi ^{2}}}\sum_{p}L_{p}^{2}%
\sum_{r=1}^{\infty }\Lambda _{p}^{-r(1+2\epsilon )}\cos (rsL_{p})=-{\frac{1}{%
2\pi ^{2}}}\Re{\frac{\partial ^{2}}{\partial s^{2}}}\sum_{p}\sum_{r=1}^{%
\infty }{\frac{1}{r^{2}}}t_{p}^{r}  \label{cor-diag}
\end{equation}
with $t_{p}=\Lambda _{p}^{-(1+2\epsilon )}e^{isL_{p}}=p^{-1-2\epsilon +is}$. 
$C_{\text{diag}}^{\epsilon }(s)$ can also be broken into two parts \cite
{Bohigas01,Bogomolny03} 
\begin{equation}
C_{\text{diag}}^{\epsilon }(s)=C_{\mathcal{\zeta }}^{\epsilon }(s)-C_{%
\mathcal{F}}^{\epsilon }(s)
\end{equation}
with 
\begin{eqnarray}
C_{\mathcal{\zeta }}^{\epsilon }(s) &=&-{\frac{1}{2\pi ^{2}}}\Re{\frac{%
\partial ^{2}}{\partial s^{2}}}\ln \mathcal{\zeta }(1+2\epsilon +is) 
\nonumber \\
C_{\mathcal{F}}^{\epsilon }(s) &=&-{\frac{1}{2\pi ^{2}}}\Re{\frac{\partial
^{2}}{\partial s^{2}}}\ln \mathcal{F}_{\epsilon }(s).
\end{eqnarray}
Here 
\begin{equation}
\ln \mathcal{F}_{\epsilon }(s)=\sum_{p}\sum_{r=1}^{\infty }\frac{r}{(r+1)^{2}%
}t_{p}^{r+1}.
\end{equation}
Alternatively, one has the following expressions by direct summation over $r$%
\begin{eqnarray}
C_{\text{diag}}^{\epsilon }(s) &=&{\frac{1}{2\pi ^{2}}}\Re\sum_{p}L_{p}^{2}%
\frac{t_{p}}{1-t_{p}},  \nonumber \\
C_{\mathcal{\zeta }}^{\epsilon }(s) &=&{\frac{1}{2\pi ^{2}}}%
\Re\sum_{p}L_{p}^{2}\frac{t_{p}}{(1-t_{p})^{2}},  \nonumber \\
C_{\mathcal{F}}^{\epsilon }(s) &=&{\frac{1}{2\pi ^{2}}}\Re\sum_{p}L_{p}^{2}%
\frac{t_{p}^{2}}{(1-t_{p})^{2}}.  \label{cor-over-p}
\end{eqnarray}
These expressions have absolute convergence. Obviously, one has $C_{\mathcal{%
F}}^{\epsilon}(s)<{\frac{1}{2}}C_{\mathcal{\zeta }}^{\epsilon}(s)$.

In order to obtain a compact form for $C_{\mathcal{\zeta }}^{\epsilon }(s)$,
we notice the following expression for the Riemann zeta function $\zeta (s)$ 
\[
\mathcal{\zeta }(s)=\frac{e^{(\ln 2\pi -1-\gamma /2)s}}{2(s-1)\Gamma (1+%
\tfrac{1}{2}s)}\prod_{n}\left( 1-\frac{s}{\gamma _{n}}\right) e^{s/\gamma
_{n}}
\]
with $\gamma _{n}=\frac{1}{2}\pm it_{n}$. One has 
\begin{equation}
{\frac{\partial ^{2}}{\partial s^{2}}}\ln \mathcal{\zeta }(1+is)={\frac{1}{%
s^{2}}}+\frac{1}{4}\psi _{1}(\tfrac{3}{2}+i\tfrac{1}{2}s){+}\sum_{n}{\frac{1%
}{(1+is-\gamma _{n})^{2}}.}  \label{deriv-zeta}
\end{equation}
Here ${\psi }_{n}(z)\equiv d^{n+1}\Gamma (z)/dz^{n+1}$ is the polygamma
function with 
\[
{\psi}_{n}(z)=(-1)^{n+1}n!\sum\limits_{j=0}^{\infty }{\frac{1}{(j+z)^{n+1}}}.
\]
Define the following function 
\begin{equation}
Z(z)\equiv {\frac{1}{(z-1)^{2}}-\sum\limits_{j=1}^{\infty }\frac{1}{%
(z+2j)^{2}}-\sum\limits_{j=1}^{\infty }\frac{1}{(z-{\frac{1}{2}}\pm
it_{j})^{2}},}
\end{equation}
one thus has 
\begin{equation}
C_{\mathcal{\zeta }}^{\epsilon }(s)={\frac{1}{2\pi ^{2}}\Re}Z(1+2\epsilon
+is).
\end{equation}

In order to obtain a compact form for the term $C_{\mathcal{F}}^{\epsilon
}(s)$, we use the following expansion for small $t$%
\begin{equation}
\sum_{q=2}^{\infty }\frac{q-1}{q^{2}}t^{q}=\sum_{n=2}^{\infty
}c_{n}\sum_{r=1}^{\infty }\frac{t^{nr}}{r}.  \label{exp-cn}
\end{equation}
The coefficients ${c_{n}}$ can be obtained by equalizing the coefficients of 
$t^{q}$ on both sides of the equation. For prime numbers and the power of 2,
one has 
\begin{eqnarray}
c_{p} &=&(p-1)/p^{2},  \nonumber \\
c_{2^{n}} &=&2^{-2n},\text{ }n\geq 1.
\end{eqnarray}
The rest of the ${c}_{n}$'s are given by the following recursive equation 
\begin{equation}
{\sum_{n\geq 2,r\geq 1,nr=q}\frac{1}{r}}c_{n}={\frac{q-1}{q^{2}}.}
\end{equation}
For example, one has ${c}_{6}=-1/18$, ${c}_{10}=-1/25$, ${c}_{12}=-1/72$, ${c%
}_{14}=-3/98$, ${c}_{15}=-8/15^{2}$.

Using the expansion (\ref{exp-cn}) and function $Z(z)$, one has 
\begin{eqnarray}
C_{\mathcal{F}}^{\epsilon }(s) &=&-{\frac{1}{2\pi ^{2}}}\Re{\frac{\partial
^{2}}{\partial s^{2}}\sum_{n=2}^{\infty }\frac{{c_{n}}}{n^{2}}}\ln \mathcal{%
\zeta }[n(1+2\epsilon +is)]  \nonumber \\
&=&{\frac{1}{2\pi ^{2}}\Re\sum_{n=2}^{\infty }n^{2}c_{n}}Z[n(1+2\epsilon
+is)].
\end{eqnarray}
One thus gets 
\begin{eqnarray}
C_{\text{diag}}^{\epsilon }(s) &=&-{\frac{1}{2\pi ^{2}}\Re\sum_{n=1}^{\infty
}n^{2}c_{n}}Z[n(1+2\epsilon +is)]  \nonumber \\
&=&-{\sum_{n=1}^{\infty }n^{2}c_{n}}C_{\mathcal{\zeta }}^{{\frac{n-1}{2}+n}%
\epsilon }(ns)  \label{cor-diag-zeta}
\end{eqnarray}
with ${c_{1}=-1}$. If one ignores the off-diagonal term $C_{\text{off}%
}^{\epsilon }(s)$, one obtains (\ref{conj-1}) with $g(z,\alpha )$ given by
Eq. (\ref{g-diag}).

The value of the correlation at the origin can also be calculated in the
following 
\[
C_{\text{diag}}^{\epsilon }(0)={\frac{1}{2\pi ^{2}}}\sum_{p}L_{p}^{2}%
\sum_{r=1}^{\infty }\Lambda _{p}^{-r(1+2\epsilon )}={\frac{1}{2\pi ^{2}}}%
\sum_{p}{\frac{L_{p}^{2}}{e^{(1+2\epsilon )L_{p}}-1}}.
\]
The average density of periodic orbits for large $L$ is $\rho
(L)=d(e^{L}/L)dL=(L-1)e^{L}/L^{2}$. Since one has 
\[
\int_{L}^{\infty }{\frac{L^{2}\rho (L)dL}{e^{(1+2\epsilon )L}-1}}\approx
\int_{L}^{\infty }e^{-2\epsilon L}LdL={\frac{1+2\epsilon L}{4\epsilon ^{2}}}%
e^{-2\epsilon L}.
\]
Simply set $L=\ln 2$, one has $C_{\text{diag}}^{\epsilon }(0)\approx
(1+2\epsilon \ln 2)/8\pi ^{2}\epsilon ^{2}4^{\epsilon }$. For example for $%
\epsilon =0.5$, one has $C_{\text{diag}}^{\epsilon }(0)\approx 0.0429$. In
the limit $\epsilon \rightarrow 0$, one has $C_{\text{diag}}^{\epsilon
}(0)\approx 1/8\pi ^{2}\epsilon ^{2}$.

\section{Numerical results}

In order to check the validity of the \emph{conjecture}, we construct $%
\delta \rho _{\epsilon }(k)$ according to Eq. (\ref{dos-2}) from a finite
sequence of $t_{n}$ such that $t_{N_{0}}\lesssim K_{0}-\Delta $ and $%
t_{N}\gtrsim K_{0}+\Delta $. Define $\delta _{N}\equiv 2\pi /\ln (t_{N}/2\pi
)$ which is the local level spacing in the neighbourhood of $t_{N}$. We use $%
\epsilon $ $\sim \delta _{N}$. One has $\delta _{N}\simeq 0.669$, $0.257$, $%
0.141$, $0.134$ for $N=10^{5}$, $10^{12}$, $10^{21}$, $10^{22}$,
respectively. The RZ we studied are obtained from Odlyzko \cite{Odlyzko}. We
find the following properties of spectral correlations of the RZ.

\subsection{Correlation invariance}

For the first $10^{5}$ Riemann \ zeros, the fluctuation part of the spectral
density $\delta \rho _{\epsilon }(k)$ is constructed. We equally divide them
into 5 segments : $[K_{n}-\Delta ,K_{n}+\Delta ]$ with $K_{n}=(2n-1)\Delta
+10$, $\Delta =7490$, and $n=1,2,\cdots 5$. Note that $t_{10^{5}}=74920.8$.
We then evaluate the auto-correlation $C_{\rho _{\epsilon }}^{r}(s)$ (\ref
{cor-2}) for each segment. We find that \emph{without any normalization, all
5 of them collapse to the same curve }. As a comparison, we also calculated
the auto-correlation for every $2\times 10^{4}$ of the first $10^{5}$ RZ.
The correlations are also the same. The relative difference between them is
less than 0.1\%.

\begin{figure}[tbp]
\center{
\includegraphics [angle=0,width=12cm]{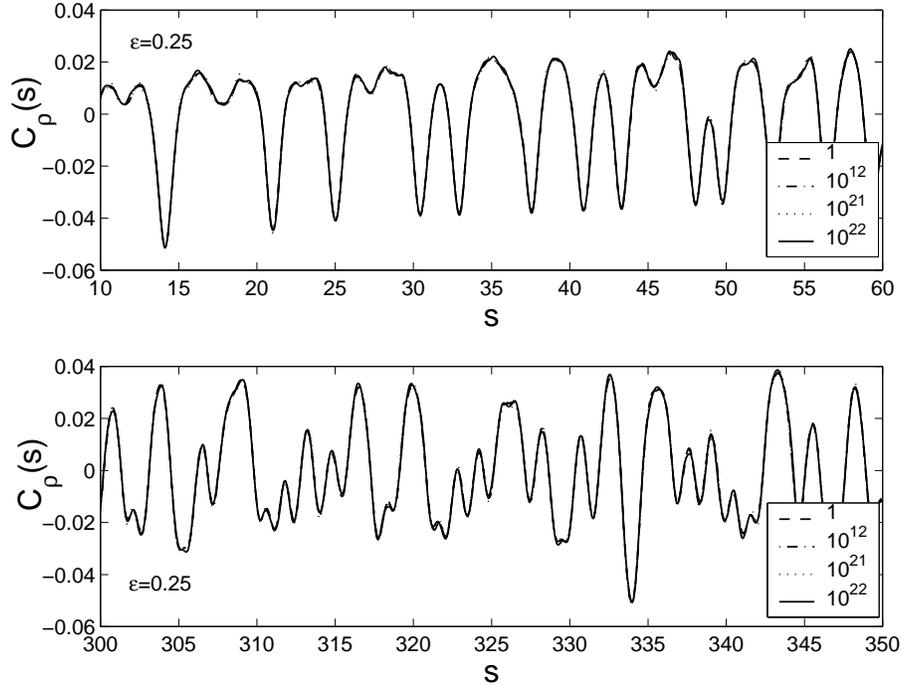}}
\caption{Invariance of spectral correlation $C_{\rho }^{r}(s;K_{0};\Delta )$
for RZ with $\epsilon =0.25$: $\{1,10^{5}\}$ (dashed), $%
\{10^{12}+1,10^{12}+10^{4}\}$ (dash-dotted), $\{10^{21}+1,10^{21}+10^{4}\}$
(dotted), and $\{10^{22}+1,10^{22}+10^{4}\}$ (solid). All curves for
different ranges are almost identical.}
\label{fig2}
\end{figure}

In the same way, we consider four groups of RZ: $\{1,10^{5}\}$, $\{10^{12}+1 
$, $10^{12}+10^{4}\}$, $\{10^{21}+1,10^{21}+10^{4}\}$, and $\{10^{22}+1$, $%
10^{22}+10^{4}\}$. For the spectral correlations of the above four segments
of RZ with the same width $\epsilon $, they also \emph{all collapse to the
same curve without any normalization }as shown in Fig.\ref{fig2}.

\begin{figure}[tbp]
\center{
\includegraphics [angle=0,width=12cm]{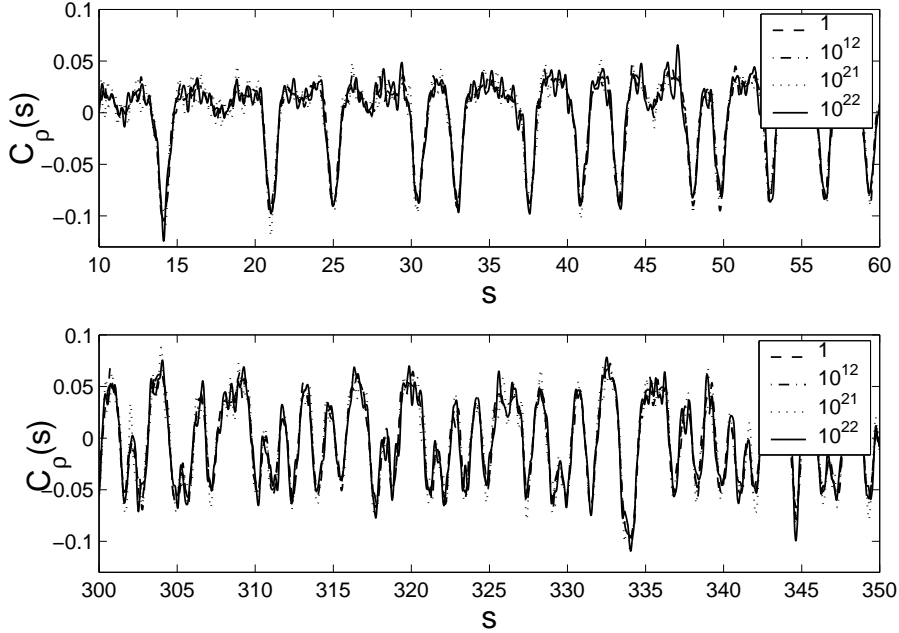}}
\caption{Same as in Fig.\ref{fig2} with $\epsilon =0.1$.}
\label{fig3}
\end{figure}

In general, one would expect that the spectral correlation will depend on
the choice of $K_{0}$ and $\Delta $ in Eq.(\ref{cor-2}). Even though the
spectrum is more congested around $t_{n}\sim K_{0}$ for larger $n$, they all
have almost the same spectral auto-correlation. We find that the main
feature of the auto-correlation is only determined by the choice of $%
\epsilon $ and independent of $K_{0}$ and $\Delta $. So no matter how high
the RZ located on the critical line, they are all correlated in the same way
as the low lying RZ. The dependence on $K_{0}$ and $\Delta $ will show up
for very small width $\epsilon $ such that $\epsilon $ $<\delta $ and very
narrow $\Delta $. This is shown in Fig.\ref{fig3} where we used $\epsilon
=0.1<\delta _{10^{22}}\simeq 0.134$.

We point out that the diagonal approximation is very accurate for the
Riemann \ zeros. Numerically calculated $C_{\rho _{\epsilon }}^{r}(s)$ are
almost indistinguishable from the direct evaluation of $C_{\text{diag}}(s)$
given in Eq. (\ref{cor-over-p}) using the first 350,000 primes for $s<500$.
For example for $\epsilon =0.25$, $0.5$, one has the standard deviation $%
\sigma =\langle C_{\text{off}}^{2}\rangle ^{1/2}/\langle C_{\rho
}^{r2}\rangle ^{1/2}\sim 0.022$, $0.031$, respectively for the spectral
correlation of the RZ $\{t_{n}\}$ with $n\sim 10^{12}$.

\subsection{Resurgence}

The invariance of the spectral correlations implies the existence of certain
intrinsic structure in the spectral auto-correlation. By \emph{resurgence} 
of RZ we mean the appearance of RZ $\{{\frac{1}{2}\pm i}t_{n}\}$ 
with small $n$ from the correlation of RZ with large $n$. 
\emph{We find that the
peaks of }$-C_{\rho _{\epsilon }}^{r}(s)$\emph{\ are located at the
positions }$s=t_{n}$.

\begin{figure}[tbp]
\center{
\includegraphics [angle=0,width=12cm]{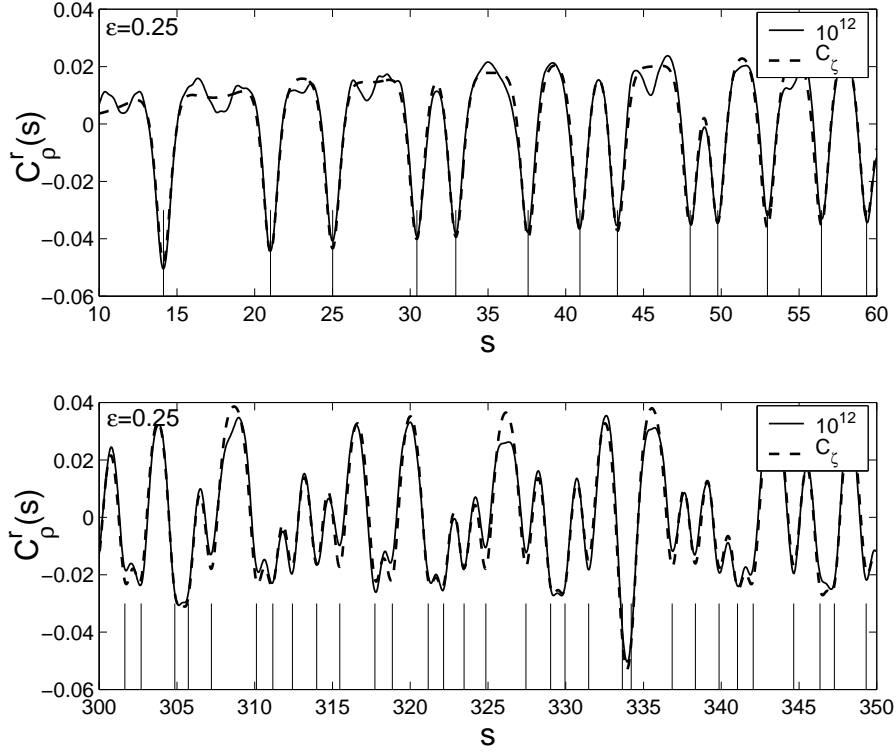}}
\caption{Resurgence of RZ in $C_{\rho }^{r}(s;K_{0},\Delta )$ for $\epsilon
=0.25$. Here $C_{\rho }^{r}(s;K_{0};\Delta )$ is the numerically calculated
auto correlation in $[K_{0}-\Delta ,K_{0}+\Delta ]$ with $%
K_{0}=(t_{10^{12}+1}+t_{10^{12}+10^{4}})/2$ and $\Delta \approx
(t_{10^{12}+10^{4}}-t_{10^{12}+1})/2$. The dashed curve is $C_{\zeta
}^{\epsilon }(s)$. Vertical lines are located at $t_{n}$ and coincide with
the valleys.}
\label{fig4}
\end{figure}

\begin{figure}[tbp]
\center{
\includegraphics [angle=0,width=12cm]{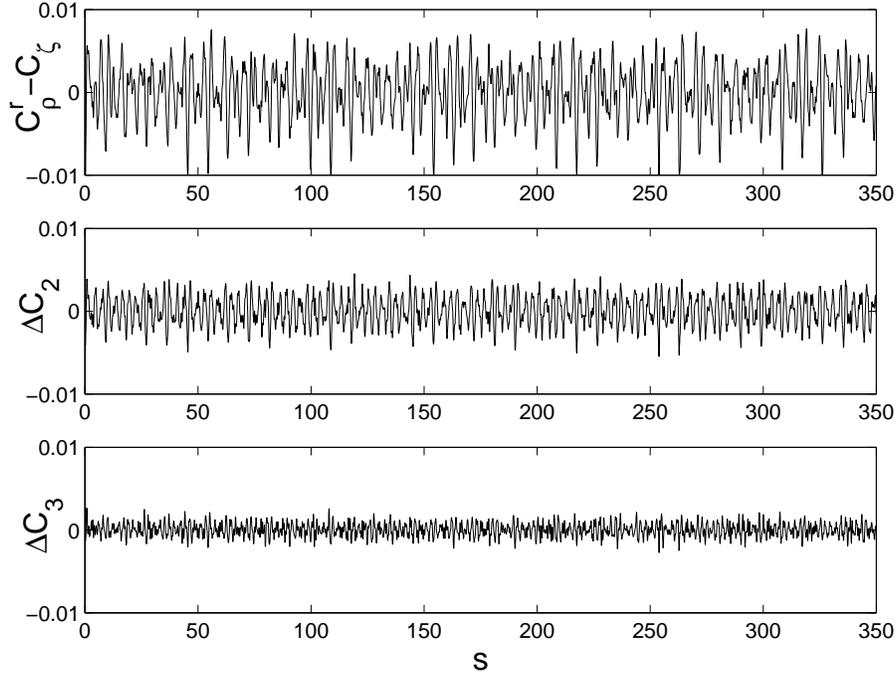}}
\caption{Resurgence of RZ in $C_{\rho }^{r}(s;K_{0},\Delta )$ for $\epsilon
=0.25$. Here $C_{\rho }^{r}(s;K_{0},\Delta )$ is the numerically calculated
auto correlation in $[K_{0}-\Delta ,K_{0}+\Delta ]$ with $K_{0}-\Delta
\approx t_{10^{12}+1}$ and $K_{0}+\Delta \approx t_{10^{12}+10^{4}}$. (a) $%
C_{\rho }^{r}(s;K_{0},\Delta )-C_{\zeta }^{\epsilon }(s)$, (b) $\Delta
C_{2}(s)$, (c) $\Delta C_{3}(s)$.}
\label{fig5}
\end{figure}

We then use our conjecture (\ref{conjecture}) to reveal the structure 
within the auto-correlation. The numerically calculated 
$C_{\rho _{\epsilon}}^{r}(s;K_{0},\Delta )$ can be approximated very well 
by $C_{\mathcal{\zeta }}^{\epsilon }(s)$ without any fitting parameters. 
This is shown in Fig. \ref{fig4} for width $\epsilon =0.25$. 
The difference $%
C_{\rho _{\epsilon }}^{r}(s)-C_{\mathcal{\zeta }}^{\epsilon }(s)$ is very
small as shown in Fig. \ref{fig5}a with $\sigma =\langle (C_{\rho _{\epsilon
}}^{r}-C_{\mathcal{\zeta }}^{\epsilon })^{2}\rangle ^{1/2}/\langle C_{\rho
}^{r2}\rangle ^{1/2}=0.174,0.155$, for $\epsilon =0.25,0.5$, respectively of
the RZ $\{t_{n}\}$ with $n\sim 10^{12}$ for $0<s<350$. 
Better approximations can be achieved by including more terms 
in $C_{\text{diag}}^{\epsilon }(s)$. Define the difference of a 
better approximation according to Eq. (\ref{cor-diag-zeta}), 
\begin{equation}
\Delta C_{2}(s)=C_{\rho _{\epsilon }}^{r}(s;K_{0},\Delta )-C_{\mathcal{\zeta 
}}^{\epsilon }(s)+C_{\rho _{\frac{1}{2}+2\epsilon }}^{r}(2s;K_{0},\Delta ),
\end{equation}
one has the standard deviation reduced to $\sigma =0.093,0.072$, for $%
\epsilon =0.25,0.5$, respectively. We further define 
\begin{equation}
\Delta C_{3}(s)=\Delta C_{2}+2C_{\rho _{1+3\epsilon }}^{r}(3s;K_{0},\Delta ),
\end{equation}
then the standard deviation $\sigma =0.041,0.036$ for $\epsilon =0.25,0.5$,
respectively. This is comparable with the off-diagonal term. The behaviors
of $C_{\rho _{\epsilon }}^{r}-C_{\mathcal{\zeta }}^{\epsilon }$, $\Delta
C_{2}$, and $\Delta C_{3}$ for $\epsilon =0.25$ are plotted in 
Fig.\ref{fig5}.

\begin{figure}[tbp]
\center{
\includegraphics [angle=0,width=12cm]{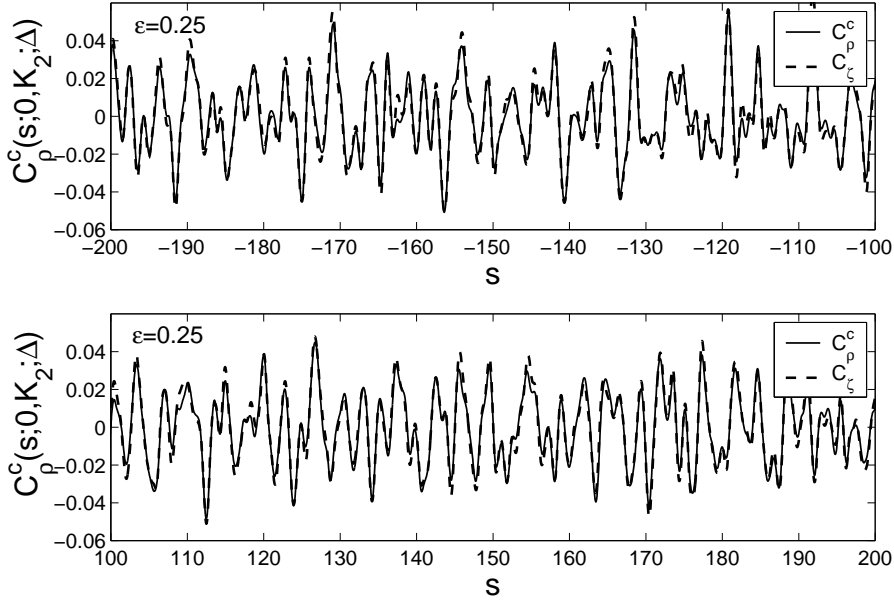}}
\caption{Resurgence of the RZ in the cross correlation $C_{\rho
}^{c}(s;0,K_{2},\Delta ))$ between $[-\Delta ,\Delta ]$ and $[K_{2}-\Delta
,K_{2}+\Delta ]$ with $K_{2}=(t_{10^{22}+1}+t_{10^{22}+10^{4}})/2$ and $%
\Delta \approx (t_{10^{22}+10^{4}}-t_{10^{22}+1})/2$. The dashed curve is $%
C_{\zeta }^{\epsilon }(K_{2}+s)$.}
\label{fig6}
\end{figure}

Resurgence can also be observed in the cross correlation $C_{\rho _{\epsilon
}}^{c}(s;K_{1},K_{2},\Delta )$ in Eq. (\ref{cor-1}) for two windows $%
[K_{1}-\Delta ,K_{1}+\Delta ]$ and $[K_{2}-\Delta ,K_{2}+\Delta ]$ with $%
K_{1}\sim 0$. Suppose we consider the set $\{t_{n}\}$ with $10^{22}<n\leq
10^{22}+10^{4}$. We use $K_{1}=0$, $%
K_{2}=(t_{10^{22}+1}+t_{10^{22}+10^{4}})/2$, and choose the window size $%
\Delta =620\sim (t_{10^{22}+10^{4}}-t_{10^{22}+1})/2$ in (\ref{cor-1}) so
that $K_{2}+\Delta \sim t_{10^{22}+10^{4}}$. The spectral function $\delta
\rho _{\epsilon }(k)$ is constructed according to Eq. (\ref{dos-1}) in
windows $[-\Delta ,\Delta ]$ and $[K_{2}-\Delta ,K_{2}+\Delta ]$. The center
distance between these two windows is $K_{2}$. The cross correlation $%
C_{\rho _{\epsilon }}^{c}(s;0,K_{2},\Delta )$ with $s\in \lbrack -\Delta
/2,\Delta /2]$ and $\epsilon =0.25$ is shown in Fig.\ref{fig6}. One can see
that $C_{\rho _{\epsilon }}^{c}(s;0,K_{2},\Delta )$ can be well approximated
by $C_{\mathcal{\zeta }}^{\epsilon }(K_{2}+s)$ with 
$\sigma=\langle(C_{\rho_{\epsilon}}^c-C_{\mathcal{\zeta}}^{\epsilon
})^{2}\rangle ^{1/2}/\langle C_{\rho }^{2}\rangle ^{1/2}\sim 0.199,0.171$
for $\epsilon =0.25,0.5$, respectively. To improve the fitting, one can use 
\begin{equation}
C_{\rho _{\epsilon }}^{c}(s;0,K_{2},\Delta )\simeq C_{\mathcal{\zeta }%
}^{\epsilon }(K_{2}+s)-C_{\rho _{{\frac{1}{2}+}2\epsilon
}}^{p}(2s;K_{2},\Delta ).
\end{equation}
Here $C_{\rho _{\epsilon }}^{p}(s;K_{2},\Delta )$ with $-\Delta \leq s\leq
\Delta $ is the cross correlation (\ref{cor-3}) between $[-K_{2}-\Delta
,-K_{2}+\Delta ]$ and $[K_{2}-\Delta ,K_{2}+\Delta ]$. This time, the
standard deviation is reduced to $\sigma =0.118,0.079$ for $\epsilon
=0.25,0.5$, respectively. Similar results are also obtained for $K_{2}\sim t_{10^{12}}$
and $K_{2}\sim t_{10^{21}}$.

\begin{figure}[tbp]
\center{
\includegraphics [angle=0,width=12cm]{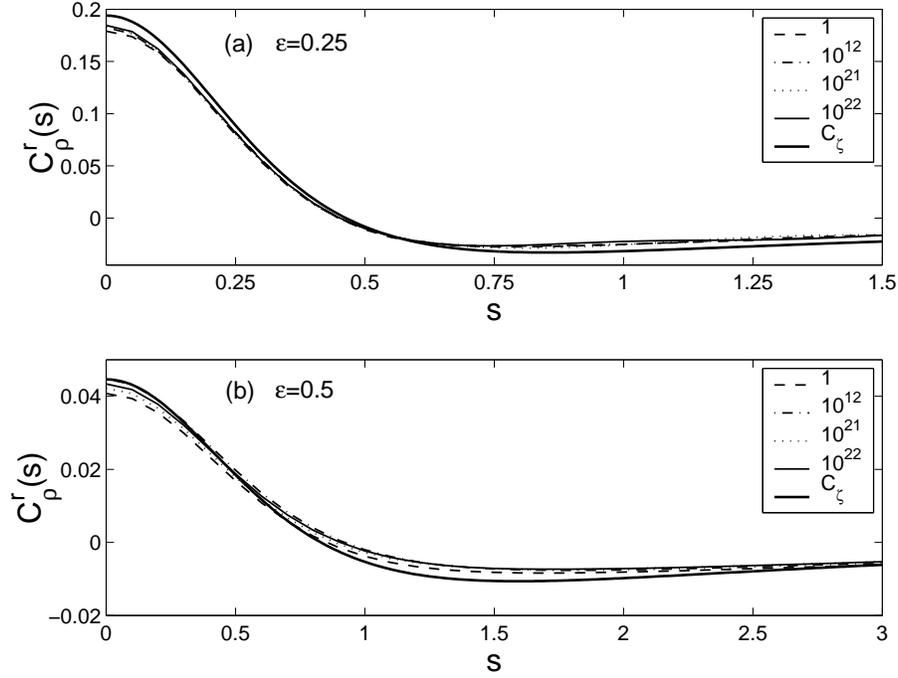}}
\caption{Small $s$ behavior of $C_{\rho }^{r}(s)$ with (a) $\epsilon =0.25$
and (b) $\epsilon =0.5$ for RZ: $\{1,10^{5}\}$ (dashed), $%
\{10^{12}+1,10^{12}+10^{4}\}$ (dash-dotted), $\{10^{21}+1,10^{21}+10^{4}\}$
(dotted), and $\{10^{22}+1,10^{22}+10^{4}\}$ (solid). The bold solid line is 
$C_{\zeta }(s)$ (Eq.(\ref{conjecture})).}
\label{fig7}
\end{figure}

We point out that the small $s$ behavior of $C_{\rho _{\epsilon }}^{r}(s)$
is not a Lorentzian, contrary to the claim in \cite{Wardlaw}. According to (%
\ref{conjecture}), one has 
\begin{equation}
C_{\rho _{\epsilon }}^{r}(s)\simeq (4\epsilon ^{2}-s^{2})/2\pi
^{2}(s^{2}+4\epsilon ^{2})^{2}  \label{cor-sm}
\end{equation}
for $s\leq 5$. So that $C_{\rho _{\epsilon }}^{r}(0)\simeq 1/(8\pi
^{2}\epsilon ^{2})$. This value is very close to that at the origin for
different auto-correlations. The behavior of $C_{\rho _{\epsilon }}^{r}(s)$
for small $s$ is shown in Fig.\ref{fig7} for different segments of RZ.

\begin{figure}[tbp]
\center{
\includegraphics [angle=0,width=12cm]{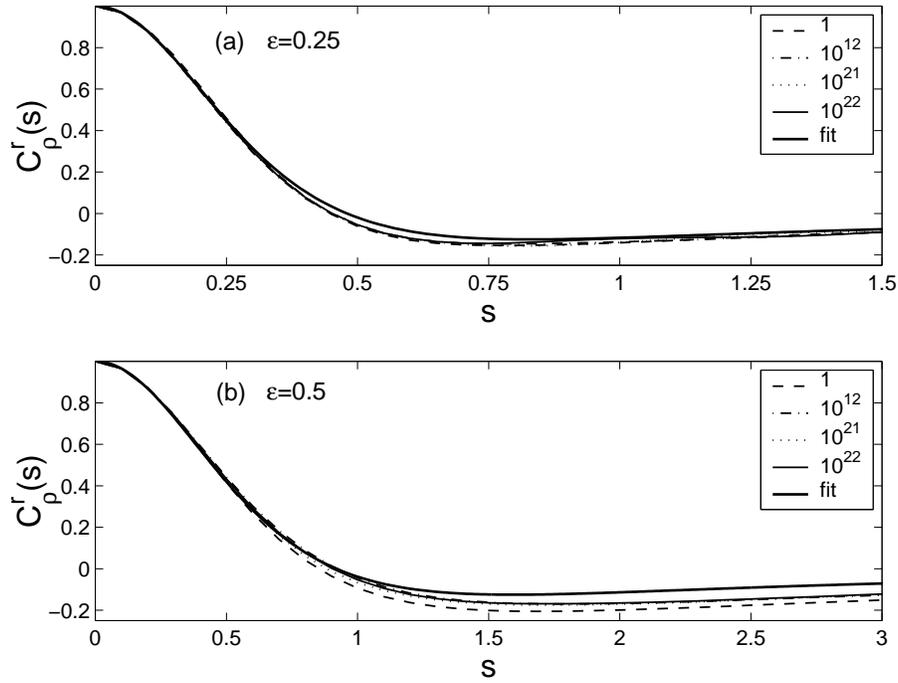}}
\caption{Small $s$ behavior of $C_{\rho }^{r}(s)/C_{\rho }^{r}(0)$ with (a) $%
\epsilon=0.25$, $\beta=0.24$ and (b) $\epsilon =0.5$ , $\beta =0.46$ for RZ: 
$\{1,10^{5}\}$ (dashed), $\{10^{12}+1,10^{12}+10^{4}\}$ (dash-dotted), $%
\{10^{21}+1,10^{21}+10^{4}\}$ (dotted), and $\{10^{22}+1,10^{22}+10^{4}\}$
(solid). Here $\beta $ is the fitting parameter in Eq. (\ref{cor-fit}).}
\label{fig8}
\end{figure}

The ratio $C_{\rho _{\epsilon }}^{r}(s)/C_{\rho _{\epsilon }}^{r}(0)$ can be
very well fitted by 
\begin{equation}
C_{\rho _{\epsilon }}^{r}(s)/C_{\rho _{\epsilon }}^{r}(0)=4\beta ^{2}(4\beta
^{2}-s^{2})/(s^{2}+4\beta ^{2})^{2}  \label{cor-fit}
\end{equation}
with fitting parameter $\beta \sim \epsilon $ for small $s$. This is shown
in Fig.\ref{fig8}. As we mentioned before, $\delta \rho _{\epsilon }(k)$ is
related to the fluctuation part of the time delay $\tau _{\text{fl}}(w)$, so 
$C_{\rho _{\epsilon }}^{r}(s)$ is related to the auto-correlation of the
time delay $c(\varepsilon ;w_{c},\Delta )$ discussed in \cite{Wardlaw} with $%
\epsilon =0.5$ and $s=2\varepsilon $. Thus for small $\varepsilon $, $%
c(\varepsilon ;w_{c},\Delta )$ can be better fitted by Eq. (\ref{cor-sm})
with $\beta =0.46$ as shown in Fig.\ref{fig8}b.

\subsection{Prophecy}

\begin{figure}[tbp]
\center{
\includegraphics [angle=0,width=12cm]{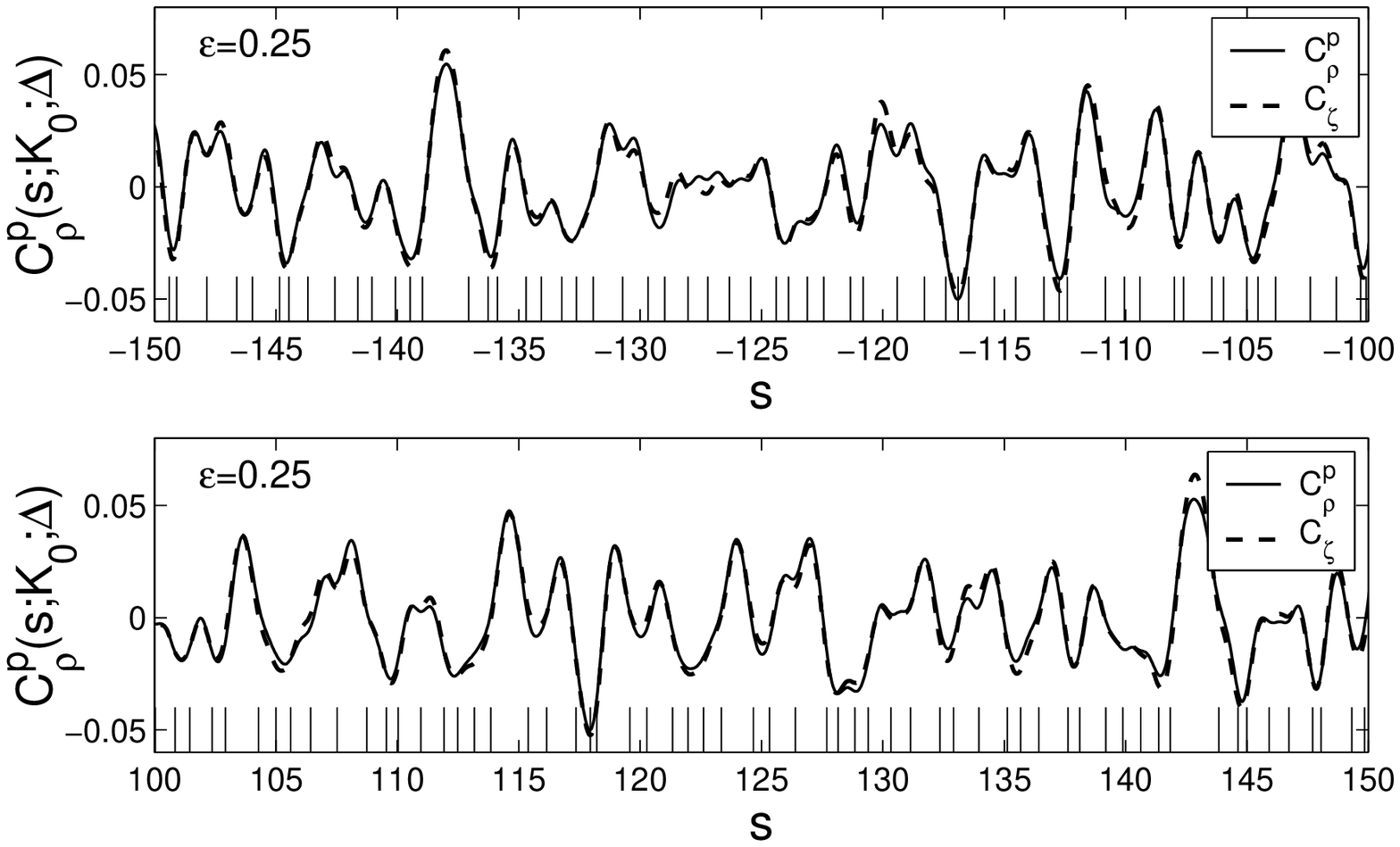}}
\caption{Prophecy of the RZ in $C_{\rho_{\epsilon}}^{p}(s;K_{0},\Delta)$ for
RZ $\{1,10^{4} \}$ with $\epsilon =0.25$, $K_{0}=4941.4$ , and $\Delta =4920$%
. The dashed curve is $C_{\zeta}^{\epsilon }(2 K_{0}+s)$. The vertical lines
are at $s=t_{n}-2 K_{0}$.}
\label{fig9}
\end{figure}

By \emph{prophecy} of RZ we mean the appearance of RZ $\{{\frac{1}{2}\pm i}%
t_{n}\}$ with large $n$ from the correlation of RZ with small $n$. We
consider the correlation $C_{\rho _{\epsilon }}^{p}(s;K_{0},\Delta )$ in Eq.
(\ref{cor-3}) for RZ $\{1,10^{4}\}$. The prophecy of new RZ from $%
C_{\rho_{\epsilon }}^{p}(s;K_{2},\Delta )$ with $\epsilon =0.25$ is shown in
Fig.\ref{fig9}. The agreement with Eq.(\ref{conj-prophecy}) is very good.

\begin{figure}[tbp]
\center{
\includegraphics [angle=0,width=12cm]{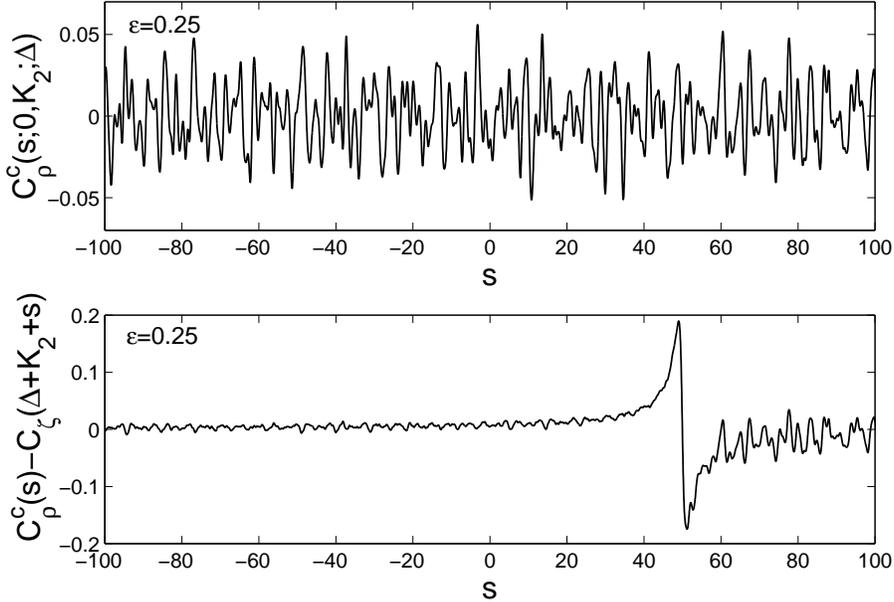}}
\caption{Prophecy of the RZ in the cross correlation $C_{\rho _{\epsilon
}}^{c}(s;\Delta ,K_{2},\Delta )$ between $[-2\Delta ,0]$ and $[K_{2}-\Delta
,K_{2}+\Delta ]$ for $\epsilon =0.25$. (a) $C_{\rho _{\epsilon
}}^{c}(s;\Delta,K_{2},\Delta )$, (b) $C_{\rho _{\epsilon }}^{c}(s;\Delta
,K_{2},\Delta )-C_{\zeta }^{\epsilon }(s+\Delta +K_{2})$. Here $C_{\zeta
}^{\epsilon }(s+ \Delta +K_{2})$ is calculated from Eq. (\ref{conjecture})
with $t_{n}$ only up to $N=10^{22}+10^{4}$.}
\label{fig10}
\end{figure}

Prophecy can also be observed in the cross correlation $C_{\rho _{\epsilon
}}^{c}(s;K_{1},K_{2},\Delta )$ in Eq. (\ref{cor-1}) for two windows $%
[K_{1}-\Delta ,K_{1}+\Delta ]$ and $[K_{2}-\Delta ,K_{2}+\Delta ]$ with $%
K_{1}+\Delta \sim 0$. As in the above subsection, we consider the set $%
\{t_{n}\}$ with $10^{22}<n\leq 10^{22}+10^{4}$. We use $K_{1}=-\Delta $, $%
K_{2}=(t_{10^{22}+1}+t_{10^{22}+10^{4}})/2$, and choose the window size $%
\Delta =(t_{10^{22}+10^{4}}-t_{10^{22}+1})/2-50=620.8$. $\delta
\rho _{\epsilon }(k)$ is calculated 
in windows $[-2\Delta,0]$ and $[K_{2}-\Delta ,K_{2}+\Delta ]$. 
The center distance between these
two windows is $K_{2}+\Delta \approx t_{10^{22}+10^{4}}$. The cross
correlation $C_{\rho _{\epsilon }}^{c}(s;\Delta ,K_{2},\Delta )$ is shown in
Fig.\ref{fig10}. If $s\leq 50$, one has resurgence of RZ since $s+\Delta
+K_{2}\leq t_{10^{22}+10^{4}}$. If $s>50$, one has prophecy of new RZ.
Similar results are also obtained for $K_{2}\sim t_{10^{12}}$ and $K_{2}\sim
t_{10^{21}}$.

The above long range correlation can be used to calculate RZ from known
ones. When applied to prophecy correlation $C_{\rho _{\epsilon }}^{p}(s)$,
new RZ can be obtained. The Fourier transformation of $C_{\rho _{\epsilon
}}(s)$ is 
\begin{equation}
\widetilde{C}(x)=\int C_{\rho _{\epsilon }}(s)e^{isx}ds\simeq -{\frac{1}{%
2\pi }x}\sum_{n}e^{it_{n}x-\gamma x}
\end{equation}
with $\gamma =1+2\epsilon $. $\widetilde{C}(x)$ will have peaks at $x=rL_{p}$%
. Since the above expression is a sum over different exponentially decaying
modes, one may use the Prony algorithm to get $t_{n}$. The above method
provides a way to calculate or at least to estimate new RZ.

\section{Discussion: Self-Duality of the Riemann spectrum}

The motivation for this work has come from experimental observations in the
microwave transmission of open $n$-disk billiards which led to the
observation of classical Ruelle-Pollicott (RP) resonances in the
auto-correlation of quantum spectra of hyperbolic $n$-disk open billiards 
\cite{Pance}. The result established a new approach to quantum-classical
correspondence by demonstrating a correspondence between the quantum and
classical resonance spectra of an open chaotic system. Applying the same
procedures developed there to the Riemann spectrum, we have arrived at the
results described in this paper. There are other parallels which are
summarized below.

\begin{enumerate}
\item  In $n$-disk open billiards mentioned above \cite{Pance}, the
experimental trace was fitted well by the sum of Lorentzians $%
T(k)=\sum_{n}b_{n}/[(k-k_{n})^{2}+k_{n}^{\prime 2}]$. Here $%
\{k_{n}+ik_{n}^{\prime }\}$ is the spectrum of eigen-wave vector resonances,
which are eigenvalues of the Helmholtz (or Laplace-Beltrami) operator, with
Dirichlet boundary conditions $(\nabla ^{2}+k^{2})\psi =0$, $\psi =0$ on $%
\partial D.$ We showed that the auto-correlation of the transmission $%
C(\kappa )=\left\langle T(k)T(k+\kappa )\right\rangle $ are well-described
as a sum of Lorentzians $C(\kappa )=\sum_{i}c_{i}/[(\kappa -\gamma
_{i}^{^{\prime }})^{2}+\gamma _{i}^{\prime \prime 2}]$. Here $\gamma
_{i}^{^{\prime }}+i\gamma _{i}^{\prime \prime }$are the \emph{classical} RP
resonances. The coefficients $c_{i}$ are related to the eigen-functions of
the Perron-Frobenius operator. The $c_{i}$ are all positive, whereas the
coefficients are negative for the RZ. For hyperbolic systems, Ruelle has
related these eigenvalues to the time-evolution of classical correlations 
\cite{Ruelle86}. \linebreak The above result can be viewed in the context of
dynamic Ruelle zeta functions which are relevant to the $n$-disk billiard.
The semiclassical Ruelle zeta-function is $\zeta _{j}(-ik)=\prod_{p}\left[
1-e^{i(kL_{p}+\pi \mu _{p}/2)}/\sqrt{\left| \Lambda _{p}\right| }\Lambda
_{p}^{j-1}\right] ^{-1}$whose poles are the quantum resonances $%
\{k_{n}+ik_{n}^{\prime }\}$, while the poles of the classical Ruelle
zeta-function $\zeta _{\beta }(s)=\prod_{p}\left( 1-e^{-sL_{p}}/\Lambda
_{p}^{\beta }\right) ^{-1}$ are the classical resonances $\{\gamma ^{\prime
}+i\gamma ^{\prime \prime }\}$ \cite{CvitanovicBook,Gaspard}. The product is
over all classical primitive periodic orbits without repetition. Here $L_{p}$
is the length of periodic orbits with the Maslov index $\mu _{p}$. $\Lambda
_{p}$ is the bigger eigenvalue of the Monodromy stability matrix for
billiard systems. Mathematically the quantum-classical correspondence can be
viewed as a duality between the poles of the semi-classical and classical
dynamic zeta functions through the auto-correlation procedure, i.e. $%
C\{k_{n}+ik_{n}^{\prime }\}\Rightarrow $ $\{\gamma _{i}^{^{\prime }}+i\gamma
_{i}^{\prime \prime }\}$.

\item  For the two-disk billiard (i.e. $n=2$), the semiclassical resonances
in wave vector space are $k_{n}=[n\pi +i(1/2)\ln \Lambda ]/(R-2a)$ with $%
\Lambda =\sigma -1+\sqrt{\sigma (\sigma -2)}$ the eigenvalue of the
instability matrix in the fundamental domain and $\sigma =R/a$. Here $n$ is
odd for $A_{1}$ representation, $n$ even and $n\neq 0$ for $A_{2}$
representation \cite{Wirzba92}. The classical RP resonances of the two-disk
system are $\gamma _{n}=[\ln \Lambda +in\pi ]/(R-2a)$. In this case there is
a direct 1-to-1 correspondence between quantum and classical resonances.

\item  In the cases of the rectangle billiard and the equal-lateral triangle
billiard, the auto-correlation $C_{\rho _{\epsilon }}^{r}(s)$ are shown \cite
{Lu2004a} to correspond to the Fourier transform of the trace of the
classical evolution operator, i.e. 
\[
C_{\rho _{\epsilon }}(s)\simeq F\Big[\mathrm{tr}\mathcal{L}%
^{t}=\sum_{p}\sum_{r=1}^{\infty }\frac{a_{p}}{rL_{p}}\delta (l-rL_{p})\Big]%
=\sum_{p}\sum_{r=1}^{\infty }\frac{a_{p}}{rL_{p}}\cos (rsL_{p}).
\]
There again the property of correlation invariance and resurgence are
clearly demonstrated.

\item  For a particle on a surface of constant negative curvature, a
self-duality exists for the quantum momentum spectrum, $\{p_{n}\}$ and the
classical spectrum $\{\gamma _{n}=\frac{1}{2}\pm ip_{n}\}$ \cite{Biswas}.
The quantum eigen-energy are given by $E_{n}=\frac{1}{4}+p_{n}^{2}$.
\end{enumerate}

Motivated by these observations, we examine a dynamic interpretation of the
Riemann zeta function. One has the following expression due to Riemann \cite
{Riemanntrace}

\begin{equation}
\sum\limits_{p}\sum_{r}{\frac{1}{r}}\delta (x-p^{r})=\frac{{1}}{{\ln x}}{-%
\frac{1}{x(x^{2}-1)\ln x}-2}\sum_{n}{\frac{\cos \left( t_{n}\ln x\right) }{%
x^{{\frac{1}{2}}}\ln x}}
\end{equation}
where $x>1$. Define $l=\ln x$, multiple $\ln x$ to both sides, one has 
\begin{equation}
\sum\limits_{p}\sum_{r}{\frac{\ln p}{p^{r}}\delta (l-r\ln p})=1-{\frac{%
e^{-2l}}{2\sinh l}-}\sum_{n}e^{-({\frac{1}{2}\pm i}t_{n})l}
\end{equation}
which is valid for $l>0$. Note the analog with the classical trace formula 
\cite{Cvitanovic}, 
\[
\text{tr}\mathcal{L}^{t}=\sum\limits_{p}\sum_{r}{\frac{L_{p}}{\Lambda
_{p}^{r}}\delta (l-r}L_{p})
\]
with $l=t\sqrt{E}$. Thus one has formally 
\begin{equation}
\text{tr}\mathcal{L}^{t}=1+\sum_{n}g_{n}e^{-\gamma
_{n}l}=1-\sum_{n=1}^{\infty }e^{-({2n+1})l}-\sum_{n=1}^{\infty }e^{-({\frac{1%
}{2}\pm i}t_{n})l}.  \label{rzclass}
\end{equation}

This is very similar to the case for the motion on the surface of constant
negative curvature, where the quantum eigen-momentum spectrum is $\{p_n\}$. 
There tr$\mathcal{L}^{t}=1+\sum_{n=1}^{\infty }e^{-({\frac{1}{2}\pm i}%
p_{n})l}$. Furthermore Biswas and Sinha have shown that the classical
resonances are given by $\gamma _{n}=\frac{1}{2}\pm ip_{n}$ \cite{Biswas} by
demonstrating that the peaks of the power spectrum of classical correlations
indeed coincide with $\{p_{n}\}$.

For the RZ, the result of performing the auto-correlation of the smoothed
density of states then leads to a self-dual ``classical''\ spectrum $\{\frac{%
1}{2}\pm it_{n}\}$. So one might say that the classical resonances are $%
\gamma _{n}={\frac{1}{2}\pm i}t_{n}$. This interpretation of RZ as RP resonances 
was suggested recently \cite{Bohigas01,Leboeuf04}.
However here $g_{n}=-1$ for the
RZ while $g_{n}=1$ for the Riemann zeta function pole and RP resonances. 
This is related to
the well-known sign problem for which a possible explanation has been
offered by Connes \cite{Connes}. .

We have clearly established a self-duality which then strongly suggests the
interpretation of the spectrum $\{t_{n}\}$ as momentum or wave vector
eigenvalues of a Helmholtz operator on a suitable domain. A similar
description has been made in which the Riemann zeta function appears in the
scattering matrix of the quantum scattering of a particle on a 2D surface of
constant negative curvature \cite{Wardlaw,levay}, so that $\{{1\over 2}t_{n}\}$ are
the real parts of the poles of the scattering matrix.

\section{Conclusion}

In this paper, we report the observation of new correlations among the RZ
that presented as a conjecture supported by analytical arguments and
numerical simulations. The spectral correlations possess three important
properties: invariance, resurgence and prophecy. The prophecy can be used to
progressively obtain new RZ from known ones. The last observation suggests
an interesting new strategy for proving the Riemann hypothesis via a
bootstrap approach starting with known zeros.

We thank F.Y. Wu and J.V. Jos\'e for discussions. This work is supported
partially by NSF-PHY-0098801.

$^{\dagger }$w.lu@neu.edu $^{\ast }$s.sridhar@neu.edu

\end{document}